\documentclass[aps,pra,twocolumn,groupedaddress,floatfix,showpacs]{revtex4}

\usepackage{graphicx}
\usepackage[latin1]{inputenc}
\usepackage{amsmath}
\usepackage{amssymb}
\usepackage{color}

\newcommand{\nc}{\newcommand}
\nc{\bra}[1]{\langle #1|}
\nc{\ket}[1]{|#1\rangle}
\nc{\braket}[1]{\left\langle #1 \right\rangle}
\nc{\equ}[1]{\begin{eqnarray*}#1\end{eqnarray*}}
\nc{\equn}[1]{\begin{eqnarray}#1\end{eqnarray}}
\nc{\dagg}{^{\dagger}}
\nc{\conj}{^{*}}
\nc{\dx}[1]{\, \mathrm{d} {#1} \,}
\nc{\Dx}[1]{\mathcal{D} {#1} \,}
\nc{\la}{\langle}
\nc{\ra}{\rangle}
\nc{\tr}{\text{tr}}
\nc{\Tr}{\text{Tr} \,}
\nc{\e}{\text{e}}
\nc{\Id}{\mathbb{1}}
\nc{\eps}{\varepsilon}
\nc{\der}[2]{\frac{\mathrm{d} {#1}}{\mathrm{d} {#2}}}
\nc{\pder}[2]{\frac{\partial {#1}}{\partial {#2}}}
\nc{\bigO}{\mathcal{O}}
\nc{\half}{\frac{1}{2}}

\nc{\Eq}[1]{Eq.~(\ref{#1})}
\nc{\eq}[1]{Eq.~(\ref{#1})}
\nc{\chap}[1]{Chapter \ref{#1}}
\nc{\Sect}[1]{Section \ref{#1}}
\nc{\sect}[1]{section \ref{#1}}
\nc{\fig}[1]{Fig.~\ref{#1}}
\nc{\Fig}[1]{Fig.~\ref{#1}}
\nc{\tabl}[1]{Table \ref{#1}}
\nc{\app}[1]{Appendix \ref{#1}}

\nc{\eg}{\emph{e.g.} }
\nc{\ie}{\emph{i.e.} }
\nc{\etal}{et al.}

\nc{\asp}[1]{#1}
\nc{\alg}[1]{#1}

\begin{document}

\title{Strong antibunching in a generalized Rabi model}

\author{Arne L. Grimsmo}\email{arne.grimsmo@ntnu.no}
\affiliation{Department of Physics, 
             The Norwegian University of Science and Technology, N-7491 Trondheim, Norway}
\affiliation{Department of Physics, 
             University of Auckland, Private Bag 92019, Auckland, New Zealand}
\author{Scott Parkins}\email{s.parkins@auckland.ac.nz}
\affiliation{Department of Physics, 
             University of Auckland, Private Bag 92019, Auckland, New Zealand}
\date{\today}

\begin{abstract}
We consider a generalized version of the Rabi model that includes a nonlinear, dispersive-type atom-field interaction in addition to the usual linear dipole coupling, as well as cavity dissipation. An effective system of this sort arises, for example, in a quantum simulation of the Rabi model based upon Raman transitions in an optical cavity QED setting [A.~L. Grimsmo and S. Parkins, Phys. Rev. A {\bf87}, 033814 (2013)]. For a suitable choice of the nonlinear interaction strength, near degeneracies of the states in the cavity-mode vacuum and single-photon subspaces, in combination with cavity loss, gives rise to an essentially closed cycle of excitations and photon emissions within these subspaces. Consequently, the cavity output field is strongly antibunched. We develop a quantum-trajectory-based description of the system that models its key properties very well, and use a simple dressed-state picture to explain the novel structure of the cavity fluorescence spectrum. We also present numerical results for a potential realization of the system using a rubidium atom coupled strongly to a high-finesse optical cavity mode.
\end{abstract}


\maketitle

\section{Introduction}

Quantum simulation is an exciting and rapidly evolving field of research, a key aim of which is to engineer evolution and interactions in well-defined and well-understood quantum systems so as to generate dynamics described by specific models of interest in specific parameter regimes \cite{Buluta09,Cirac12}. These dynamics are typically not able to be realized naturally in such ``clean'' form, and may offer exotic and useful physical behaviors and properties, as well as potential insights into more fundamental phenomena to which they may be relevant.

Engineered systems may also enable generalizations of an original target model that simply have no parallel in conventional, ``naturally occurring'' systems, and thereby open up possibilities for novel or unforeseen behavior. In recent work \cite{Grimsmo13}, we proposed a scheme based on interactions in optical cavity quantum electrodynamics (cavity QED) for the simulation of qubit-oscillator dynamics in the ultrastrong-coupling regime; that is, for the realization of a system described by the Rabi Hamiltonian ($\hbar=1$),
\begin{align}\label{eq:H_Rabi}
  H_{\rm R} = \frac{\omega_0}{2} \sigma_z + \omega a\dagg a + g(\sigma_+ + \sigma_-)(a+a\dagg) ,
\end{align}
operating in a regime where the effective qubit and oscillator frequencies, $\omega_0$ and $\omega$, respectively, are comparable in magnitude to, or even less than, the coupling strength, $g$.
Here, $\{\sigma_z,\sigma_\pm=\frac{1}{2}(\sigma_x\pm i\sigma_y)\}$ are two-state (Pauli) operators for the qubit, and $a$ ($a\dagg$) is the annihilation (creation) operator for the quantized oscillator. In the proposed scheme, the qubit states are two stable (hyperfine) ground states of an atom, while the oscillator is simply the cavity field mode. Raman transitions between the atomic states are driven by the cavity mode and auxiliary laser fields, such that $g$ is given by a Raman transition rate, while dispersive energy shifts due to the laser fields and atom-cavity dipole coupling determine the effective frequencies $\omega_0$ and $\omega$. In this way, $\{\omega_0,\omega,g\}$ can be brought to the same magnitude.

However, the scheme of \cite{Grimsmo13} also gives rise to a nonlinear coupling of the form
\begin{align}\label{eq:H_NL}
H_{\rm NL} = \frac{U}{2} \sigma_z a\dagg a , 
\end{align}
where the coupling strength $U$ is also determined by a dispersive energy shift due to the atom-cavity dipole coupling and can in practice be chosen to match or even exceed the magnitudes of $\omega_0$, $\omega$ and $g$. The form (\ref{eq:H_NL}) is normally recognized as an effective Hamiltonian for the dispersive limit of (\ref{eq:H_Rabi}); i.e., if $|\omega_0-\omega|\gg g$, then the atom-cavity interaction of (\ref{eq:H_Rabi}) can effectively be reduced to (\ref{eq:H_NL}) (with $U\propto g^2/(\omega_0-\omega)$ and hence $|U|\ll g$). 

The scheme of \cite{Grimsmo13} allows $H_{\rm R}$ and $H_{\rm NL}$ to exist {\em simultaneously} and on an {\em equal footing} in the description of the cavity QED system. We therefore have an example of an engineered quantum system that has no direct parallel in the conventional description of atom-cavity interactions. 

That this can give rise to very interesting and novel behavior was illustrated briefly in \cite{Grimsmo13}, where the regime $\omega_0=\omega$ with $|U|\sim g\sim \{\omega_0,\omega\}$ was considered with the addition of cavity dissipation. In particular, critical-type behavior was observed in the mean photon number $\langle a^\dagger a\rangle$, atomic inversion $\langle\sigma_z\rangle$, and intensity correlation function $g^{(2)}(0)=\langle a^\dagger a^\dagger aa\rangle/\langle a^\dagger a\rangle^2$ around the values $U=\pm 2\omega$. The significance of these two values follows straightforwardly from the form that the total Hamiltonian, $H=H_{\rm R}+H_{\rm NL}$,  takes at these values, i.e.,
\begin{align}
H_{U=\pm 2\omega} =& \frac{\omega_0}{2} \sigma_z + \omega (1\pm\sigma_z)  a\dagg a \nonumber
\\
& ~~~ + g(\sigma_+ + \sigma_-)(a+a\dagg).
\end{align}
In the limit $g\rightarrow 0$, the eigenstates of this Hamiltonian are the bare states $|n,g\rangle$ and $|n,e\rangle$, where $n$ denotes the cavity photon number and $\{ g,e\}$ the atomic state. More importantly, at $U=2\omega$ the eigenstates $|n,g\rangle$ ($n=0,1,2,\ldots$) are degenerate, with energy $-\omega_0/2$, while at $U=-2\omega$ the eigenstates $|n,e\rangle$ ($n=0,1,2,\ldots$) are degenerate, with energy $\omega_0/2$. For finite $g$, this (infinite) degeneracy results in pronounced excitation of the system at $U=\pm 2\omega$, mitigated however by cavity dissipation to give a finite, dynamical steady state.

\begin{figure}[t!]
  \includegraphics[scale=0.6]{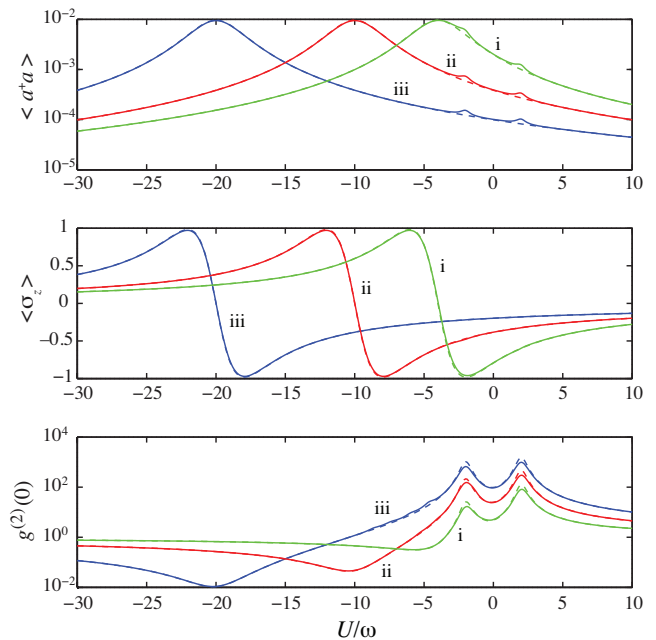}
  \caption{\label{fig:ada_sz_g2} (Color online). Mean intracavity photon number $\langle a^\dagger a\rangle$ (top), atomic inversion $\langle\sigma_z\rangle$ (middle), and intensity correlation function $g^{(2)}(0)$ (bottom) as a function of nonlinear coupling strength $U$ for $g/\omega=0.1$, $\kappa/\omega=0.2$, and (i) $\omega_0/\omega=2$ (green), (ii) $\omega_0/\omega=5$ (red), and (iii) $\omega_0/\omega=10$ (blue). Solid lines are numerical solutions of the master equation (\ref{eq:ME}), while dashed lines are approximate results (\ref{eq:ada_th}), (\ref{eq:sz_th}), and (\ref{eq:g2_th}) from the quantum trajectory analysis.}
\end{figure}

In this paper, however, we focus instead on the behavior around $U=-2\omega_0$, with comparatively small $\omega$. For this choice of $U$, the qubit states are effectively ``flipped'' between the zero- and single-photon subspaces, i.e., 
\begin{align}
H_{U=-2\omega_0} =& \frac{\omega_0}{2}\left( 1 - 2a^\dagger a \right) \sigma_z + \omega  a\dagg a \nonumber
\\
& ~~~ + g(\sigma_+ + \sigma_-)(a+a\dagg).
\end{align}
The significance of this exchange is illustrated in Fig.~\ref{fig:ada_sz_g2}, where we plot $\langle a^\dagger a\rangle$, $\langle\sigma_z\rangle$, and $g^{(2)}(0)$ as a function of the nonlinear coupling strength $U$ for several values of the ratio $\omega_0/\omega$, computed from steady state solutions of the master equation for the density operator $\rho$,
\begin{align}\label{eq:ME}
\frac{d\rho}{dt} = -i[H,\rho ] + \kappa (2a\rho a^\dagger - a^\dagger a\rho - \rho a^\dagger a ) ,
\end{align}
where $\kappa$ is the cavity field decay rate.
One sees distinct resonances in the photon number centered at $U=-2\omega_0$, while the atomic inversion undergoes a sharp ``flip'' around this point and attains a maximum (minimum) value close to $+1$ ($-1$) at $U=-2\omega_0-2\omega$ ($U=-2\omega_0+2\omega$) for the parameters chosen. Perhaps most notably though, a broad region of strong photon antibunching, $g^{(2)}(0)\ll 1$, occurs about these values, with a minimum in $g^{(2)}(0)$ at $U=-2\omega_0$. The minimum value of $g^{(2)}(0)$ decreases rapidly as the ratio $\omega_0/\omega$ increases. Strong photon bunching is still observed around the values $U=\pm 2\omega$, as seen in \cite{Grimsmo13}, but otherwise the dominant features in the operator expectation values are those associated with the degeneracies at $U=-2\omega_0\pm 2\omega$. 

Modeling, interpreting, and quantifying the behavior of this system around the point $U=-2\omega_0$ is the main purpose of this paper. Briefly, this behavior is related to degeneracies occurring (for $g\rightarrow 0$) at $U=-2\omega_0\pm 2\omega$. In particular, at $U=-2\omega_0+2\omega$ the states $|0,e\rangle$ and $|1,g\rangle$ have the same energy, $\omega_0/2$, while at $U=-2\omega_0-2\omega$ the states $|0,g\rangle$ and $|1,e\rangle$ both have energy $-\omega_0/2$. As we will see, these degeneracies (or near-degeneracies, as is the case for $U=-2\omega_0$ with $\omega_0\gg\omega$ and $g\neq 0$) together with the linear atom-field coupling and cavity photon emissions enable a novel closed cycle within the zero- and single-photon subspaces, while excitation of higher-photon-number states is far off resonance and thereby suppressed, the obvious result of which is strong photon antibunching.

This mechanism for producing antibunched light is novel in the way that it derives from the energy level structure imposed by the nonlinear coupling, $H_{\rm NL}$, and from the form of the atom-field coupling in the Rabi Hamiltonian, $H_{\rm R}$, i.e., it depends equally on the ``rotating'' ($a^\dagger\sigma_-$) and ``counter-rotating'' ($a^\dagger\sigma_+$) terms. It complements so-called ``photon blockade'' mechanisms based upon the nonlinear structure of the Jaynes-Cummings model with strong coupling and resonant (laser) excitation of a vacuum Rabi resonance \cite{Tian92,Birnbaum05,Lang11} or upon dynamical regulation conditioned on the state of a two-level atom \cite{Dayan08}. 

Photon antibunching in the ultrastrong coupling regime of cavity QED has recently also been studied in \cite{Ridolfo12}. However, in that work, a system described by (\ref{eq:H_Rabi}) was subjected to an additional coherent driving field, and coupled to a reservoir with a different structure than we consider here. No non-linear coupling of the form (\ref{eq:H_NL}) was present, and the mechanism for antibunching was fundamentally the same as in \cite{Tian92,Birnbaum05}. Their modeling of the reservoir is relevant to recent experimental realizations of systems in the ultrastrong coupling regime based upon circuit QED or semiconductor microcavities \cite{Gunter09}, whereas we consider a ``conventional'' treatment of the reservoir, relevant to cavity QED realizations such as that considered in \cite{Grimsmo13}.

\section{Weak excitation limit: quantum trajectory analysis}

For small $g$ the cavity mode is only weakly excited (as evidenced by the small photon numbers in Fig.~\ref{fig:ada_sz_g2}), suggesting that an expansion of the system state in terms of just the lowest few Fock states of the cavity mode should suffice. Furthermore, one can assume that at any particular \asp{instant in time} the state of the system is essentially pure, and that collapses of the state due to, in our case, photon emissions from the cavity mode are infrequent. For our particular system, as described by (\ref{eq:ME}), one finds that there are in fact two orthogonal manifolds of states that the system may occupy at a given time; in particular, in between photon emissions the approximate state of the system may be described by one or other of the following two forms:
\begin{widetext}
\begin{align}
\ket{\psi^{(1)}(t)} &= \e^{-i\omega_0 t /2} \left( \alpha^{(1)}(t)\ket{0,e} + \beta^{(1)}(t)\ket{1,g} + \mu^{(1)}(t)\ket{2,e} +  \mathcal{O}(g^3) \right) ,
\\
\ket{\psi^{(2)}(t)} &= \e^{+i\omega_0 t /2} \left( \alpha^{(2)}(t)\ket{0,g} + \beta^{(2)}(t)\ket{1,e} +  \mu^{(2)}(t)\ket{2,g} + \mathcal{O}(g^3) \right) ,
\end{align}
\end{widetext}
where it is assumed that $|\alpha^{(k)}(t)|^2\simeq 1\gg |\beta^{(k)}(t)|^2\gg |\mu^{(k)}(t)|^2$ ($k=1,2$). Note that a photon emission (corresponding to a state collapse described by $a\ket{\psi^{(k)}(t)}$) switches the system state from one manifold to the other, and that the probability of a photon emission in the time interval $(t,t+dt)$ is given by $2\kappa |\beta^{(k)}(t)|^2dt$. The evolution of the state in between collapses is given by 
\begin{align}
\frac{d}{dt} \ket{\psi^{(k)}(t)} = -iH_{\rm eff} \ket{\psi^{(k)}(t)} ~~~(k=1,2)
\end{align}
where $H_{\rm eff}$ is the non-Hermitian effective Hamiltonian
\begin{align}
H_{\rm eff} = H - i\kappa a^\dagger a .
\end{align}
The equations of motion for the amplitudes in the pure state expansion are then (setting $\alpha^{(1,2)}(t)=1$)
\begin{align}
\dot{\beta}^{(1)} &= -ig -i(\omega -\omega_0-U/2 - i\kappa )\beta^{(1)} - i\sqrt{2} g\mu^{(1)} ,
\\
\dot{\mu}^{(1)} &=-i\sqrt{2} g\beta^{(1)} - i(2\omega + U - 2i\kappa )\mu^{(1)} ,
\end{align}
and
\begin{align}
\dot{\beta}^{(2)} &= -ig -i(\omega +\omega_0+U/2 - i\kappa )\beta^{(2)} - i\sqrt{2} g\mu^{(2)} ,
\\
\dot{\mu}^{(2)} &=-i\sqrt{2} g\beta^{(2)} - i(2\omega - U - 2i\kappa )\mu^{(2)} ,
\end{align}
which yield, to leading order in $g$, the (quasi-)steady state values
\begin{align}
\beta_{\rm ss}^{(1)} &\simeq -\frac{g}{\omega-\omega_0-U/2-i\kappa} , \\
\mu_{\rm ss}^{(1)} &\simeq \frac{g^2}{\sqrt{2}(\omega+U/2-i\kappa)(\omega - \omega_0 - U/2-i\kappa)},
\end{align}
and
\begin{align}
\beta_{\rm ss}^{(2)} &\simeq -\frac{g}{\omega+\omega_0+U/2-i\kappa} , \\
\mu_{\rm ss}^{(2)} &\simeq \frac{g^2}{\sqrt{2}(\omega-U/2-i\kappa)(\omega + \omega_0 + U/2-i\kappa)}.
\end{align}

\asp{Averaged over time, the state of the system will be a mixed state, approximated by a steady state density operator of the form}
\begin{align}\label{eq:rho_ss}
\rho_{\rm ss} = p_1\ket{\psi_{\rm ss}^{(1)}}\bra{\psi_{\rm ss}^{(1)}} + p_2\ket{\psi_{\rm ss}^{(2)}}\bra{\psi_{\rm ss}^{(2)}},
\end{align}
where  $p_1+p_2=1$ and
\begin{align}
\frac{p_1}{p_2} = \frac{|\beta_{\rm ss}^{(2)}|^2}{|\beta_{\rm ss}^{(1)}|^2}.
\end{align}
This ratio simply reflects the relative stability of the two manifolds with respect to photon emission. We can also write
\begin{align}\label{eq:p1p2}
p_1 = \frac{\xi}{1+\xi}, \qquad p_2 = \frac{1}{1+\xi},
\end{align}
where
\begin{align}
\xi \equiv \frac{|\beta_{\rm ss}^{(2)}|^2}{|\beta_{\rm ss}^{(1)}|^2} = \frac{\left[\omega-\left(\omega_0 + U/2\right)\right]^2 + \kappa^2}{\left[\omega+\left(\omega_0 + U/2\right)\right]^2 + \kappa^2}.
\end{align}

\subsection{Steady state expectation values}

From (\ref{eq:rho_ss}), we may deduce approximate expressions for steady state values of the photon number, atomic inversion, and intensity correlation function as follows:
\begin{align}
\braket{a\dagg a} &\simeq p_1 |\beta_{\rm ss}^{(1)}|^2 + p_2 |\beta_{\rm ss}^{(2)}|^2 \nonumber \\
&= \frac{g^2}{(\omega_0+U/2)^2+\omega^2+\kappa^2} ,
\label{eq:ada_th}
\end{align}
\begin{align}
\braket{\sigma_z} &\simeq p_1 \left( 1 - |\beta_{\rm ss}^{(1)}|^2\right) - p_2 \left( 1 - |\beta_{\rm ss}^{(2)}|^2\right) 
\nonumber \\
&= -\frac{2\omega (\omega_0+U/2)}{(\omega_0+U/2)^2+\omega^2+\kappa^2} ,
\label{eq:sz_th}
\end{align}
and
\begin{widetext}
\begin{align}
g^{(2)}(0) &\simeq 2\frac{p_1|\mu_{\rm ss}^{(1)}|^2 + p_2|\mu_{\rm ss}^{(2)}|^2}{\left(p_1|\beta_{\rm ss}^{(1)}|^2 + p_2|\beta_{\rm ss}^{(2)}|^2\right)^2} 
\nonumber \\
&= \frac{1}{2} \left( \omega^2 +(\omega_0+U/2)^2 +\kappa^2 \right) \left( \frac{1}{(\omega+U/2)^2+\kappa^2} + \frac{1}{(\omega-U/2)^2+\kappa^2}  \right) .
\label{eq:g2_th}
\end{align}
\end{widetext}

These expressions are also plotted in Fig.~\ref{fig:ada_sz_g2} and show excellent agreement with the ``exact'' numerical results over virtually all of the range of $U/\omega$ considered. The only significant discrepancy occurs around the degeneracy points at $U=\pm 2\omega$, where, given the nature of the degeneracy, one might expect that the truncated basis of the trajectory approach is inadequate.

From the approximate analytical expressions, we see explicitly that:
\begin{enumerate}
\item
 The mean photon number takes the form of a Lorentzian in $U$, with a maximum value of $g^2/(\omega^2+\kappa^2)$ at $U=-2\omega_0$ and a half-width of $2\sqrt{\omega^2+\kappa^2}$.
\item
The atomic inversion crosses zero at $U=-2\omega_0$ and attains extremum values at $U=-2\omega_0\pm 2\omega$ of
\begin{align}
\braket{\sigma_z}\left|_{U=-2\omega_0\pm 2\omega}\right. = \mp \frac{2\omega^2}{2\omega^2+\kappa^2} ,
\end{align}
which, for $\omega\gg\kappa$, approach $\mp 1$.

\item
The intensity correlation function has a minimum value at $U=-2\omega_0$ given by
\begin{widetext}
\begin{align}
g^{(2)}(0) \left|_{U=-2\omega_0} \right. &= \frac{1}{2} \left( \frac{\omega^2+\kappa^2}{(\omega-\omega_0)^2+\kappa^2} + \frac{\omega^2+\kappa^2}{(\omega+\omega_0)^2+\kappa^2} \right)
\\
&\simeq \frac{\omega^2+\kappa^2}{\omega_0^2+\kappa^2} ~~~ \textrm{for} ~~~ \omega_0\gg\omega .
\end{align}
\end{widetext}
If, in addition, one has $\omega_0\gg\omega\gg\kappa$, then $g^{(2)}(0) \left|_{U=-2\omega_0} \right. \simeq \omega^2/\omega_0^2\ll 1$ and light emitted from the cavity mode is strongly antibunched, as illustrated in Fig.~\ref{fig:ada_sz_g2}.

\end{enumerate}

\begin{figure}[h!]
  \includegraphics[scale=0.55]{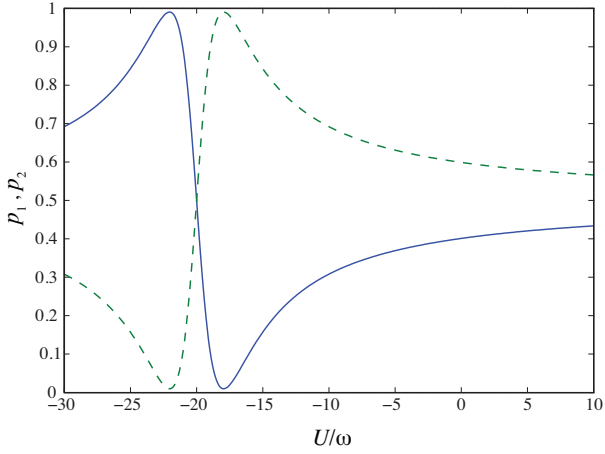}
  \caption{\label{fig:p1p2} (Color online). Probabilities $p_1$ (solid) and $p_2$ (dashed) of being in the states $\ket{\psi^{(1)}}$ and $\ket{\psi^{(2)}}$, respectively, as a function of nonlinear coupling strength $U$ for $g/\omega=0.1$, $\kappa/\omega=0.2$, and $\omega_0/\omega=10$.}
\end{figure}

\begin{figure}
  \includegraphics[scale=0.5]{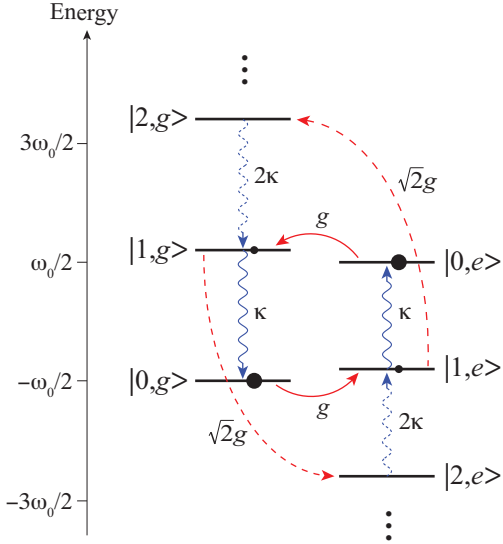}
  \caption{\label{fig:levels} (Color online). Energy level diagram in the weak excitation ($g\ll\omega_0,\omega$) limit for $U=-2\omega_0$, with $\omega_0/\omega=10$.}
\end{figure}

It is also interesting to observe the variation in the populations $p_1$ and $p_2$ of the two orthogonal manifolds as a function of $U$, as given by (\ref{eq:p1p2}). This is shown in Fig.~\ref{fig:p1p2}. At the degeneracy points $U=-2\omega_0\pm2\omega$ we observe an extreme asymmetry in the populations, with $p_2\simeq 1$ or $p_1\simeq 1$ ($\xi=\kappa^2/(4\omega^2+\kappa^2)\ll 1$ or $\xi=(4\omega^2+\kappa^2)/\kappa^2\gg 1$), respectively. At $U=-2\omega_0$, however, we have $p_1=p_2$, and one can put forward a simple physical picture, depicted in Fig.~\ref{fig:levels}, of the dynamics of the system leading to antibunching in the photon statistics. In particular, to a good approximation the system executes a closed cycle of excitations and photon emissions amongst the states involving just the vacuum or one photon, i.e., $\ket{0,g}\rightarrow\ket{1,e}\leadsto\ket{0,e}\rightarrow\ket{1,g}\leadsto\ket{0,g}$. Excitations to states of higher photon number are suppressed by the combination of small $g$ and large energy gap ($\sim 2\omega_0$ for $\ket{1,e}\rightarrow\ket{2,g}$ and $\ket{1,g}\rightarrow\ket{2,e}$), hence the strong degree of antibunching observed at and around the value $U=-2\omega_0$. 
We emphasize that the nonlinear coupling, $H_{\rm NL}$, is essential to achieving this variation on the photon blockade effect by bringing the states $\ket{0,g}$ and $\ket{1,e}$ (and similarly $\ket{0,e}$ and $\ket{1,g}$) close to resonance with eachother, while keeping the states $\ket{2,g}$ and $\ket{2,e}$ far from resonance.

\subsection{Intensity correlation function $g^{(2)}(\tau)$}

\begin{figure}
  \includegraphics[scale=0.48]{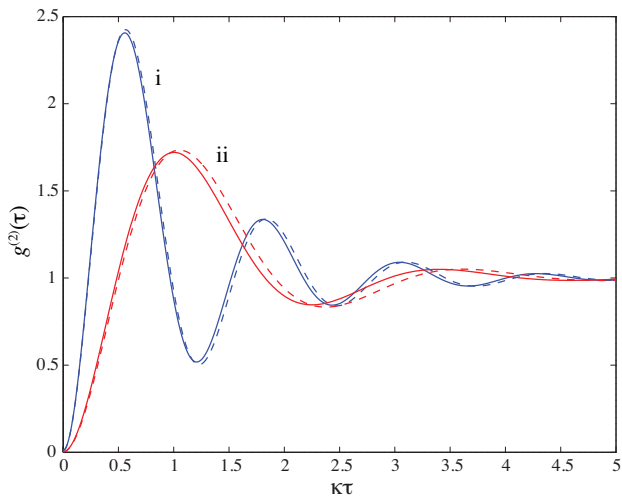}
  \caption{\label{fig:g2tau_1} (Color online). Intensity correlation function $g^{(2)}(\tau)$ for $U=-2\omega_0$ and $\{\omega_0,g\}/\kappa=\{ 50,0.5\}$, with (i) $\omega/\kappa=5$ (blue) and (ii) $\omega/\kappa=2.5$ (red). Solid lines are numerical solutions of the master equation (\ref{eq:ME}), computed using the quantum regression theorem, while dashed lines are approximate analytical results from the quantum trajectory analysis. }
\end{figure}

The time-dependent intensity correlation function, $g^{(2)}(\tau)=\braket{a^\dagger a^\dagger (\tau)a(\tau)a}/\braket{a^\dagger a}^2$, which gives the conditional probability for a photon emission at a time $t=\tau$ given an emission at $t=0$, can also be estimated from the trajectory analysis. Consider the effect of a photon emission on the steady state density operator $\rho_{\rm ss}$. The state is projected onto a new state described by the (normalized) density operator
\begin{align}
\tilde{\rho}(0) &= \frac{a\rho_{\rm ss}a^\dagger}{{\rm Tr}\{a\rho_{\rm ss}a^\dagger\}} \nonumber
\\
&= p_1\ket{\tilde{\psi}^{(2)}(0)}\bra{\tilde{\psi}^{(2)}(0)} + p_2\ket{\tilde{\psi}^{(1)}(0)}\bra{\tilde{\psi}^{(1)}(0)}  ,
\end{align}
where we have defined
\begin{align}
\ket{\tilde{\psi}^{(1)}(0)} &= \frac{\beta_{\rm ss}^{(2)}\ket{0,e}+\sqrt{2}\mu_{\rm ss}^{(2)}\ket{1,g}}{\sqrt{p_1 |\beta_{\rm ss}^{(1)}|^2 + p_2 |\beta_{\rm ss}^{(2)}|^2}} \nonumber
\\
&\equiv \tilde{\alpha}^{(1)}(0)\ket{0,e} + \tilde{\beta}^{(1)}(0)\ket{1,g} ,
\label{eq:tilde01}
\\
\ket{\tilde{\psi}^{(2)}(0)} &= \frac{\beta_{\rm ss}^{(1)}\ket{0,g}+\sqrt{2}\mu_{\rm ss}^{(1)}\ket{1,e}}{\sqrt{p_1 |\beta_{\rm ss}^{(1)}|^2 + p_2 |\beta_{\rm ss}^{(2)}|^2}} \nonumber
\\
&\equiv \tilde{\alpha}^{(2)}(0)\ket{0,g} + \tilde{\beta}^{(2)}(0)\ket{1,e} ,
\label{eq:tilde02}
\end{align}
and assumed that $|\beta_{\rm ss}^{(1,2)}|^2\gg |\mu_{\rm ss}^{(1,2)}|^2$. The correlation function can then be approximated by the expression
\begin{align}\label{eq:g2tau_theory}
g^{(2)}(\tau) \simeq \frac{p_2|\tilde{\beta}^{(1)}(\tau)|^2+p_1|\tilde{\beta}^{(2)}(\tau)|^2}{p_1|\beta_{\rm ss}^{(1)}|^2 + p_2|\beta_{\rm ss}^{(2)}|^2} ,
\end{align}
where $\tilde{\beta}^{(1)}(\tau)$ and $\tilde{\beta}^{(2)}(\tau)$ are solutions of the amplitude equations of motion derived for the trajectory model, but with initial conditions $\tilde{\alpha}^{(1,2)}(0)$ and $\tilde{\beta}^{(1,2)}(0)$ as defined in (\ref{eq:tilde01}) and (\ref{eq:tilde02}). For sufficiently small $g$ we can, as before, make the simplification $\tilde{\alpha}^{(1,2)}(\tau)\simeq\tilde{\alpha}^{(1,2)}(0)$ and derive the solutions
\begin{widetext}
\begin{align}\label{eq:tildabeta1}
\tilde{\beta}^{(1)}(\tau) &= \frac{\sqrt{2}\mu_{\rm ss}^{(2)}}{\sqrt{p_1|\beta_{\rm ss}^{(1)}|^2 + p_2|\beta_{\rm ss}^{(2)}|^2}} {\rm e}^{-i(\omega-\omega_0-U/2-i\kappa)\tau} - \frac{\beta_{\rm ss}^{(2)}\beta_{\rm ss}^{(1)}}{\sqrt{p_1|\beta_{\rm ss}^{(1)}|^2 + p_2|\beta_{\rm ss}^{(2)}|^2}} \left[ {\rm e}^{-i(\omega-\omega_0-U/2-i\kappa)\tau} - 1\right] ,
\\
\tilde{\beta}^{(2)}(\tau) &= \frac{\sqrt{2}\mu_{\rm ss}^{(1)}}{\sqrt{p_1|\beta_{\rm ss}^{(1)}|^2 + p_2|\beta_{\rm ss}^{(2)}|^2}} {\rm e}^{-i(\omega+\omega_0+U/2-i\kappa)\tau} - \frac{\beta_{\rm ss}^{(1)}\beta_{\rm ss}^{(2)}}{\sqrt{p_1|\beta_{\rm ss}^{(1)}|^2 + p_2|\beta_{\rm ss}^{(2)}|^2}} \left[ {\rm e}^{-i(\omega+\omega_0+U/2-i\kappa)\tau} - 1\right] .
\label{eq:tildabeta2}
\end{align}
\end{widetext}
A comparison of (\ref{eq:g2tau_theory}) (using (\ref{eq:tildabeta1}) and (\ref{eq:tildabeta2}) as above) with numerical results is shown in Fig.~\ref{fig:g2tau_1} for the case $U=-2\omega_0$ with a couple of values of $\omega$. The agreement, particularly at shorter times, is very good and improves further as the ratio $g/\omega$ is decreased. As predicted by the approximate theoretical expression and illustrated in Fig.~\ref{fig:g2tau_1}, the initial rise time of $g^{(2)}(\tau)$ is determined in this limit by the oscillator frequency $\omega$.

\begin{figure}
  \includegraphics[scale=0.48]{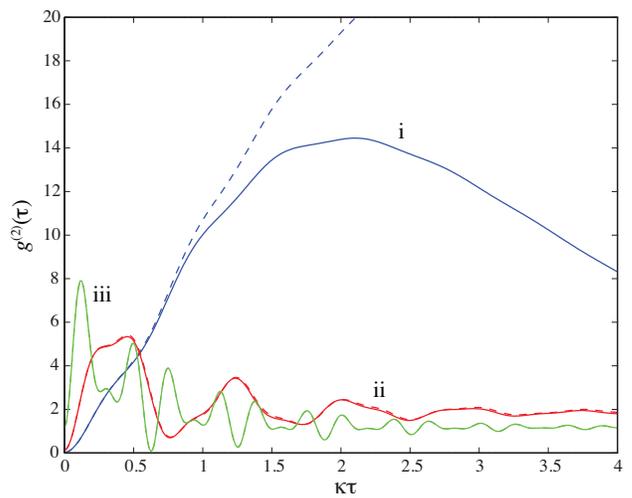}
  \caption{\label{fig:g2tau_3} (Color online). Intensity correlation function $g^{(2)}(\tau)$ for $\{\omega,\omega_0,g\}/\kappa=\{ 5,50,0.5\}$, with (i) $U=-2\omega_0+2\omega$ (blue), (ii) $U=-2\omega_0+4\omega$ (red), and (iii) $U=-2\omega_0+10\omega$ (green). Solid lines are numerical solutions of the master equation (\ref{eq:ME}), computed using the quantum regression theorem, while dashed lines are approximate analytical results from the quantum trajectory analysis in the weak excitation (small $g$) limit.}
\end{figure}

Away from the value $U=-2\omega_0$ the approximate analytic expression given above generally describes the behavior quite well, but at (and in the vicinity of) the degeneracy points $U=-2\omega_0\pm2\omega$ it works only for sufficiently short times and fails badly at longer times, as shown in Fig.~\ref{fig:g2tau_3}. Here, the first photon emission takes the system very far away from its steady state and the coherent coupling $g$ plays a far more significant role in the subsequent evolution. Much better agreement for the case $U=-2\omega_0+2\omega$ on the timescale considered in Fig.~\ref{fig:g2tau_3} is obtained by using more accurate solutions of the coupled amplitude equations (i.e., by allowing for $\tilde{\alpha}^{(1,2)}(\tau)\neq\tilde{\alpha}^{(1,2)}(0)$). Nevertheless, the weak-excitation theory does correctly predict, for $U=-2\omega_0+2\omega$, a rise time of $g^{(2)}(\tau)$ on the order of $\kappa^{-1}$.

\section{Cavity emission spectrum}

\begin{figure}
  \includegraphics[scale=0.48]{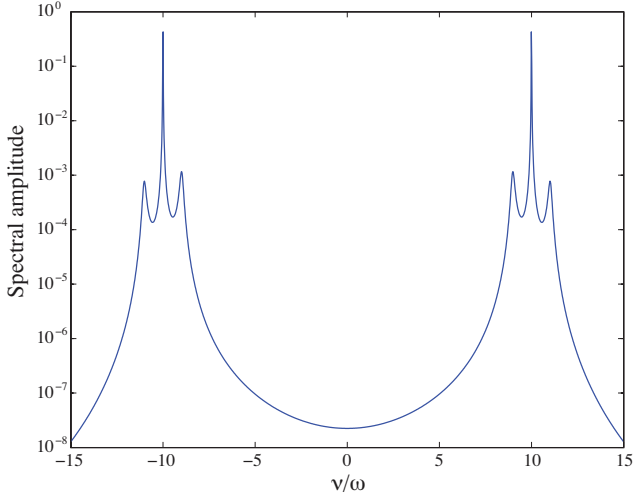}
  \caption{\label{fig:spec} (Color online). Cavity emission spectrum for $U=-2\omega_0$, with $\omega_0/\omega=10$, $g/\omega=0.1$, and $\kappa/\omega=0.1$.}
\end{figure}

\begin{figure}
  \includegraphics[scale=0.55]{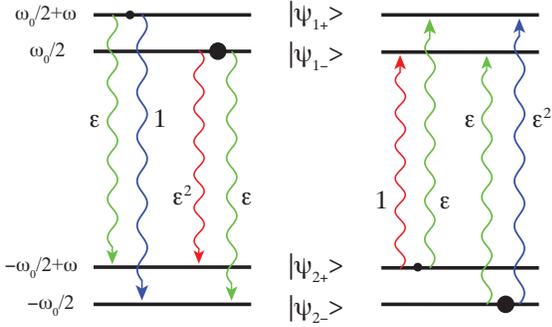}
  \caption{\label{fig:dressedstates1} (Color online). Dressed state picture of the emission processes for $U=-2\omega_0$ and small $g$.  Shown next to each of the emission channels is the absolute value of the transition amplitude $\bra{\psi_{2\pm}}a\ket{\psi_{1\pm}}$ or $\bra{\psi_{1\pm}}a\ket{\psi_{2\pm}}$. The approximate energies of the dressed states are shown on the left.}
\end{figure}

Given the interesting dynamics apparent from the discussion above, it is also worth examining the power spectrum of the light emitted by the cavity mode, which is defined by
\begin{align}
S(\nu) = \int_{-\infty}^\infty d\tau\; C(\tau) {\rm e}^{-i\nu\tau} ,
\end{align}
with $C(\tau)=\braket{a^\dagger(\tau)a} - \braket{a^\dagger}\braket{a}$ (note, however, that the mean cavity field amplitude for our system is in fact always zero, i.e., $\braket{a}=0$). This spectrum is shown in Fig.~\ref{fig:spec} for the case $U=-2\omega_0$ and small $g/\omega$. Sharp, dominant peaks appear in the spectrum at the frequencies $\nu=\pm\omega_0$, but distinctive sidebands also feature at frequencies $\nu=\pm\omega_0\pm\omega$. Better understanding of the structure of the spectrum requires consideration of the dressed states of the system, as illustrated in Fig.~\ref{fig:dressedstates1} (where only the most relevant states are shown). The (unnormalized) dressed states take the general forms $\ket{\psi_{1+}}=\ket{1,g}+\epsilon\ket{0,e}$, $\ket{\psi_{1-}}=\ket{0,e}-\epsilon\ket{1,g}$, $\ket{\psi_{2+}}=\ket{1,e}+\epsilon\ket{0,g}$, and $\ket{\psi_{2-}}=\ket{0,g}-\epsilon\ket{1,e}$, with $\epsilon\sim g/\omega\ll 1$. Most of the population in the system is equally distributed between the states $\ket{\psi_{1-}}$ and $\ket{\psi_{2-}}$, between which transitions occur at frequency $\omega_0$ at a rate proportional to $\kappa\epsilon^2\sim \kappa (g/\omega)^2$, hence the narrow linewidth of the spectral  peaks at $\nu=\pm\omega_0$. The sidebands to these peaks result primarily from the transitions $\ket{\psi_{1+}}\leadsto\ket{\psi_{2-}}$ and $\ket{\psi_{2+}}\leadsto\ket{\psi_{1-}}$ and have a linewidth (FWHM) close to $2\kappa$.

\begin{figure}
  \includegraphics[scale=0.48]{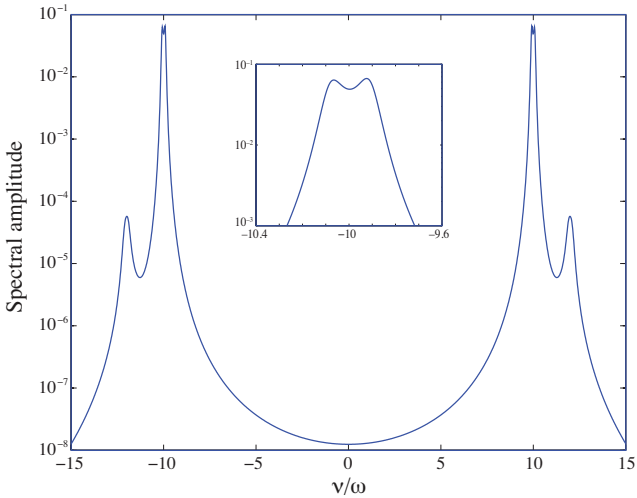}
  \caption{\label{fig:spec3} (Color online). Cavity emission spectrum for $U=-2\omega_0-2\omega$, with $\omega_0/\omega=10$, $g/\omega=0.1$, and $\kappa/\omega=0.1$. The inset shows a close up of the spectrum around $\nu=-\omega_0$.}
\end{figure}

\begin{figure}
  \includegraphics[scale=0.55]{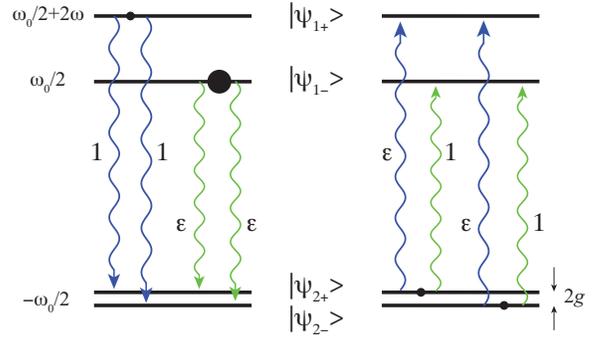}
  \caption{\label{fig:dressedstates2} (Color online). Dressed state picture of the emission processes for $U=-2\omega_0-2\omega$ and small $g$. Shown next to each of the emission channels is the absolute value of the transition amplitude $\bra{\psi_{2\pm}}a\ket{\psi_{1\pm}}$ or $\bra{\psi_{1\pm}}a\ket{\psi_{2\pm}}$. The approximate energies of the dressed states are shown on the left.}
\end{figure}

In contrast, the spectrum for $U=-2\omega_0-2\omega$, shown in Fig.~\ref{fig:spec3}, exhibits only a single sideband at frequencies $\nu\simeq -\omega_0-2\omega$ and $\nu\simeq \omega_0+2\omega$, while the dominant peaks centered at $\nu=\pm\omega_0$ display a slight splitting of $\sim 2g$. The same splitting is not resolved in the sidebands with the parameters of Fig.~\ref{fig:spec3}, but appears for smaller values of $\kappa$. The relevant (unnormalized) dressed states, depicted in Fig.~\ref{fig:dressedstates2}, are now $\ket{\psi_{1+}}=\ket{1,g}+\epsilon\ket{0,e}$ and $\ket{\psi_{1-}}=\ket{0,e}-\epsilon\ket{1,g}$, with $\epsilon\sim g/(2\omega)\ll 1$, and $\ket{\psi_{2\pm}}=\ket{0,g}\pm\ket{1,e}$. The bulk of the system population presides in the state $\ket{\psi_{1-}}$, and transitions between this state and the lower doublet of $\ket{\psi_{2+}}$ and $\ket{\psi_{2-}}$ (split by $\sim 2g$) gives rise to the dominant spectral feature around $\nu=\pm\omega_0$. The sideband results primarily from the transitions $\ket{\psi_{1+}}\leadsto\ket{\psi_{2\pm}}$.

\section{Stronger excitation}

\begin{figure}
  \includegraphics[scale=0.6]{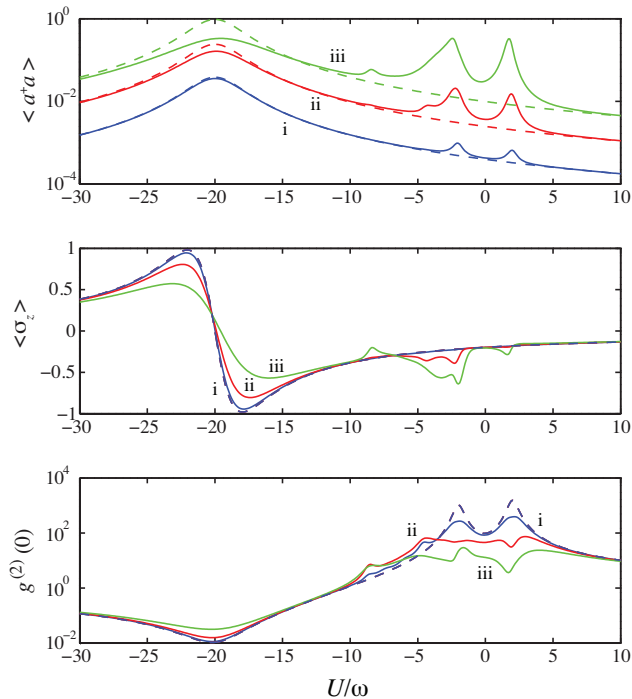}
  \caption{\label{fig:ada_sz_g2_g} (Color online). Mean intracavity photon number $\langle a^\dagger a\rangle$ (top), atomic inversion $\langle\sigma_z\rangle$ (middle), and intensity correlation function $g^{(2)}(0)$ (bottom) as a function of nonlinear coupling strength $U$ for $\omega_0/\omega=10$, $\kappa/\omega=0.2$, and (i) $g/\omega=0.2$ (blue), (ii) $g/\omega=0.5$ (red), and (iii) $g/\omega=1.0$ (green). Solid lines are numerical solutions of the master equation (\ref{eq:ME}), while dashed lines are approximate analytical results (\ref{eq:ada_th}), (\ref{eq:sz_th}), and (\ref{eq:g2_th}) from the quantum trajectory analysis. Note that for $\langle\sigma_z\rangle$ and $g^{(2)}(0)$ the analytical result is independent of $g$.}
  \end{figure}

With larger values of $g$ the system is more strongly excited and the approximate analysis of the previous section starts to deviate more significantly from the numerical solutions of the master equation. This is highlighted in Fig.~\ref{fig:ada_sz_g2_g}, where we see increasing discrepancies, in particular  around the degeneracy points. For small $|U|/\omega$, the behavior of the expectation values becomes especially complicated.

\begin{figure}
  \includegraphics[scale=0.55]{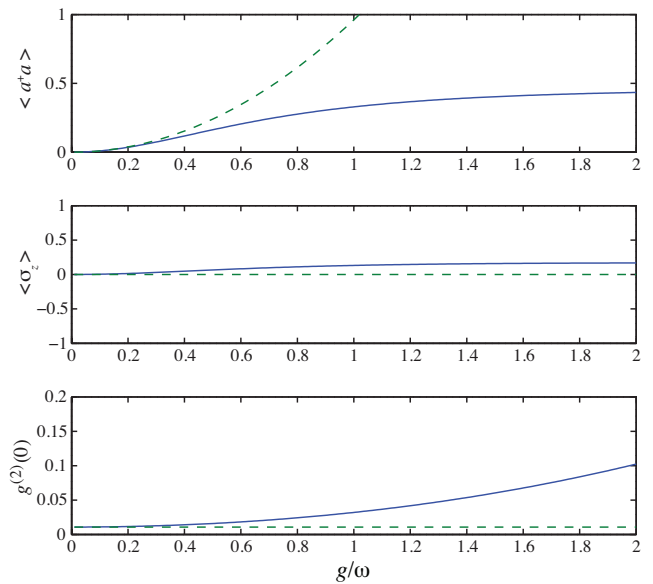}
  \caption{\label{fig:ada_sz_g2_saturation} (Color online).  Mean intracavity photon number $\langle a^\dagger a\rangle$ (top), atomic inversion $\langle\sigma_z\rangle$ (middle), and intensity correlation function $g^{(2)}(0)$ (bottom) as a function of linear coupling strength $g$ for $\omega_0/\omega=10$, $\kappa/\omega=0.2$, and $U=-2\omega_0$. Solid lines are numerical solutions of the master equation (\ref{eq:ME}), while dashed lines are approximate analytical results (\ref{eq:ada_th}), (\ref{eq:sz_th}), and (\ref{eq:g2_th}) from the quantum trajectory analysis in the weak excitation limit.}
\end{figure}

It is interesting though to focus again on the point $U=-2\omega_0$, which we do in Fig.~\ref{fig:ada_sz_g2_saturation} by plotting expectation values, obtained numerically from the full master equation, as a function of $g$. The results of the weak-excitation theory are also plotted and show reasonable agreement with the full model for values of $g/\omega\lesssim 0.4$. For larger $g$ the mean photon number starts to saturate towards a value of 0.5 as populations of the zero- and one-photon states equalize. Interestingly, the degree of antibunching remains large, with, for example, $g^{(2)}(0)\simeq 0.1$ even for $g/\omega=2$ (where $\braket{a^\dagger a}\gtrsim 0.4$).

\begin{figure}
  \includegraphics[scale=0.5]{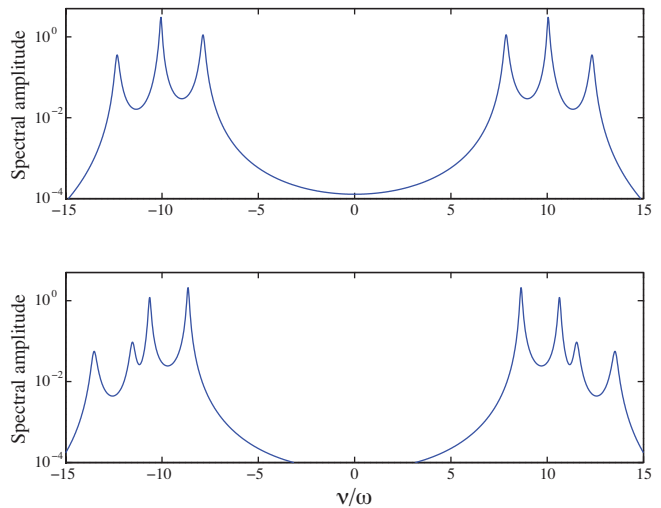}
  \caption{\label{fig:spec-large-g} (Color online). Cavity emission spectrum for $U=-2\omega_0$ (top) and $U=-2\omega_0-2\omega$ (bottom), with $\omega_0/\omega=10$, $g/\omega=1$, and $\kappa/\omega=0.1$. }
\end{figure}

For larger $g$ the cavity emission spectrum for $U=-2\omega_0$ retains the same general structure, but the sidebands grow in strength relative to the central peaks and shift significantly in frequency, as shown in Fig.~\ref{fig:spec-large-g}. For $U=-2\omega_0-2\omega$, the splitting of the states in the lower doublet ($\ket{\psi_{2\pm}}$) by $2g$ becomes clearly manifest in the spectrum and in fact plays a dominant role in determining the location of the main resonances (Fig.~\ref{fig:spec-large-g}).

\section{Practical realization}\label{sect:realization}

\begin{figure}
  \includegraphics[scale=0.4]{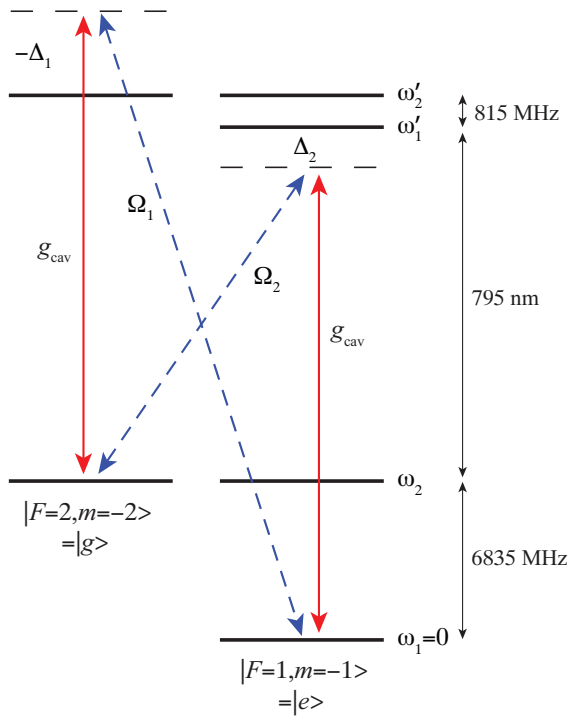}
  \caption{\label{fig:Rb87scheme} (Color online). Scheme for simulation of the generalized Rabi model, as proposed in \cite{Grimsmo13}. Shown are the relevant levels of the D1 line in ${}^{87}$Rb, together with the cavity ($g_{\rm cav}$) and laser ($\Omega_{1,2}$) fields. }
\end{figure}

\begin{figure}
  \includegraphics[scale=0.52]{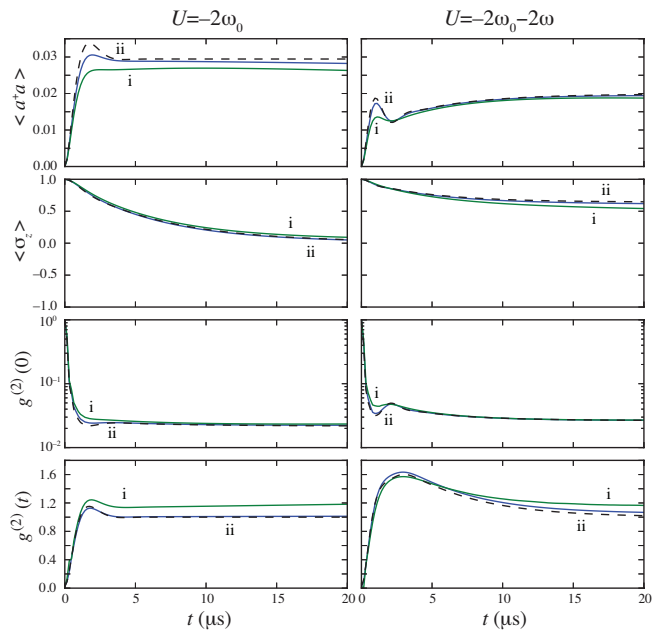}
  \caption{\label{fig:Rb87results} (Color online). Time evolution of the expectation values $\braket{a^\dagger a}$, $\braket{\sigma_z}$, and $g^{(2)}(0)$. Initially the cavity mode is in the vacuum state and the atom is in the state $\ket{e}$. Dashed lines are numerical solutions of the master equation (\ref{eq:ME}) for parameters $\{\omega_0,\omega,g,\kappa\}/(2\pi)=\{2.0,0.2,0.05,0.2\}~{\rm MHz}$ with $U/(2\pi)=-4.0~{\rm MHz}$ (left column) and $U/(2\pi)=-4.4~{\rm MHz}$ (right column). Solid lines are numerical solutions of the master equation involving the full ${}^{87}$Rb level structure with (i) $g_{\rm cav}/(2\pi)=50$~MHz and (ii) $g_{\rm cav}/(2\pi)=100$~MHz, and other parameters as given in Table \ref{tabl:Rb87params}. For this model the time-dependent intensity correlation function, $g^{(2)}(t)=\braket{a^\dagger a^\dagger (t)a(t)a}/\braket{a^\dagger a}^2$, shown in the last row, was computed using the quantum regression theorem with the density operator at $t=30~\mu {\rm s}$ taken as the initial (``stationary'') state (while for the idealized model the steady state density operator was used).}
\end{figure}

As mentioned in the introduction, a proposal for simulating the generalized Rabi model has been put forward in \cite{Grimsmo13}. This proposal is based upon cavity-plus-laser-mediated Raman transitions between ground hyperfine states of an alkali atom, as depicted in Fig.~\ref{fig:Rb87scheme}. Under suitable conditions the dynamics of the system is accurately described by the master equation (\ref{eq:ME}), with $\omega_0$, $\omega$, and $U$ determined by dispersive light shifts and $g$ by a Raman transition rate, while the effective two-state atom is formed by the ground state hyperfine levels $\ket{F=2,m=-2}$ and $\ket{F=1,m=-1}$. Example sets of (optical) cavity and laser parameters giving values of $\{\omega_0,\omega,U,g\}$ in the MHz range are shown in Table \ref{tabl:Rb87params} and were used to obtain the numerical results shown in Fig.~\ref{fig:Rb87results} for simulations involving the full atomic structure; $g_{\rm cav}$ is the atom-cavity dipole coupling strength \cite{gcav}, $\Omega_{1,2}$ are the laser Rabi frequencies, and $\Delta_{1,2}$ and $\delta_{\rm cav}$ are detunings defined by
\begin{align}
\Delta_1 = \omega_2^\prime -\omega_{L_1} , ~~~ \Delta_2 = \omega_1^\prime - (\omega_{L_1}+\omega_{L_2})/2 ,
\end{align}
\begin{align}
\delta_{\rm cav} = \omega_{\rm cav} - (\omega_{L_1}+\omega_{L_2})/2,
\end{align}
where $\omega_{L1}$, $\omega_{L2}$, and $\omega_{\rm cav}$ are frequencies of the laser fields and cavity mode, respectively. \asp{Note that the scheme of \cite{Grimsmo13} requires that $|\Delta_{1,2}|\gg \{ g_{\rm cav},|\Omega_{1,2}|,|\delta_{\rm cav}|\}$.}
The numerical values we choose for $g_{\rm cav}$ and the cavity field decay rate, $\kappa$, correspond to the strong coupling regime of optical cavity QED ($g_{\rm cav}^2/\kappa\gamma\gg 1$, where $\gamma$ is the spontaneous emission rate for the excited state of ${}^{87}$Rb), as should be feasible, for example, with microtoroidal resonators and an atom coupled to the evanescent field of a high-finesse whispering gallery mode \cite{Spillane05,Aoki06}.

\begin{table}
  \begin{tabular}{c}
  \begin{tabular}{c c c c c c c}
    \multicolumn{7}{c}{Cavity and laser parameters} \\
    \hline
    Set & $g_{\rm cav}$ & $\Omega_1$ & $\Omega_2$ & $\Delta_1$ & $\Delta_2$ & $\delta_{\rm cav}$ \\
    \hline
    $U=-4.0$ & 100 & -9.16 & -1.10 & -7478.2 & -1459.7 &  -2.25 \\
                  ~ & 50 & -17.1 & -0.673 & -6993.0 & -974.5 &  -1.92 \\
    $U=-4.4$ & 100 & -9.09 & -1.01 & -7420.6 & -1402.1 &  -2.45 \\
                  ~ & 50 & -17.1 & -0.618 & -6978.3 & -959.8 &  -2.12 \\
    \hline
 \end{tabular}
   \end{tabular}
  \caption{The parameter sets for the data of Fig.~\ref{fig:Rb87results} computed using the full ${}^{87}$Rb model, as described in Section \ref{sect:realization}. All numbers are in units of $(2\pi)\cdot$MHz. These sets give effective parameters $\omega_0/(2\pi) = 2.0$ MHz,  $\omega/(2\pi) = 0.2$ MHz, $g/(2\pi) = 0.05$ MHz, and $U/(2\pi) = -4.0$~MHz or  $-4.4$~MHz.}
  \label{tabl:Rb87params}
\end{table}

The results shown in Fig.~\ref{fig:Rb87results} for the time evolution of the expectation values and for the intensity correlation function $g^{(2)}(t)$ demonstrate good agreement, particularly for the larger value of $g_{\rm cav}$ considered, between the effective model, Eq.~(\ref{eq:ME}), and the realization based on Raman transitions in a ${}^{87}$Rb atom (computed from a master equation incorporating the full level structure of the D1 line in ${}^{87}$Rb). Importantly, the possibility of achieving the key feature of strong antibunching, $g^{(2)}(0)\ll 1$, at $U=-2\omega_0$ is clearly illustrated. Notable differences in behavior between $U=-2\omega_0$ and $U=-2\omega_0-2\omega$, particularly with regards to timescales for the approach to the steady state, are also evident.

\section{Conclusion}

In conclusion, we have presented and analyzed novel behavior of a generalized (nonlinear) and dissipative version of the Rabi model. For suitable choices of the nonlinear interaction strength, degeneracies or near-degeneracies arise between pairs of states in the vacuum and single-photon subspaces, enabling a novel cycle of excitations and photon emissions almost solely within these two subspaces. As a consequence, the cavity output field is strongly antibunched, while the cavity emission spectrum exhibits a simple, intuitive resonance structure. 

Given the richness of the behavior of the generalized Rabi model as considered here and in \cite{Grimsmo13}, it would be interesting to explore other possibilities, perhaps in circuit QED or cavity optomechanics, for physically realizing a system that is also described by this model. A collective-$N$-atom version of the current work is also interesting to consider \alg{\cite{Dimer07,Grimsmo13b}}, with the possibility of additional degeneracies and concomitant novel behavior. For example, \asp{with small $N$} we have found, under similar operating conditions to those considered here, that strong photon antibunching occurs at $U=-2N\omega_0$ when $N$ is odd, whereas strong photon bunching occurs at $U=-N\omega_0$ when $N$ is even.

\begin{acknowledgments}
  ALG is grateful for the hospitality shown at the University of Auckland when the present work was in progress. The authors thank Howard Carmichael for discussions and comments on the paper, and acknowledge the contribution of NeSI high-performance computing facilities to the results of this research. These facilities are provided by the New Zealand eScience Infrastructure and funded jointly by NeSI's collaborator institutions and through the Ministry of Science \& Innovation's Research Infrastructure Programme. URL https://www.nesi.org.nz.
\end{acknowledgments}


\begin{thebibliography}{25}
\bibitem[]{gcav}
Note that $g_{\rm cav}$ as specified in Table \ref{tabl:Rb87params} actually gives the single-photon coupling strength for the $\ket{F=2,m=2}\leftrightarrow\ket{F'=3,m'=3}$ cycling transition on the D2 line of ${}^{87}$Rb. The effective single-photon coupling strength is therefore less than this for the transitions employed in the scheme of Fig.~\ref{fig:Rb87scheme}.
\end{thebibliography}
\end{document}